\documentclass[a4paper,12pt]{article}
\usepackage[top=60pt,bottom=60pt,left=70pt,right=70pt]{geometry}
\usepackage{amsmath, amssymb, graphicx, hyperref}
\hypersetup{
    colorlinks=false,
    pdfborder={0 0 0}
}
\usepackage[greek,english]{babel}
\usepackage{makecell}
\usepackage{authblk}
\usepackage{subcaption}
\usepackage{siunitx}
\usepackage{lineno}
\usepackage{setspace} 

\usepackage[export]{adjustbox}
\usepackage{rotating}
\usepackage[table]{xcolor}

\usepackage{xr}

\makeatletter
\newcommand*{\addFileDependency}[1]{
\typeout{(#1)}
\@addtofilelist{#1}

\IfFileExists{#1}{}{\typeout{No file #1.}}
}\makeatother
\newcommand*{\myexternaldocument}[1]{%
\externaldocument{#1}%
\addFileDependency{#1.tex}%
\addFileDependency{#1.aux}%
}
\myexternaldocument{si}

\author[a]{Julian Konrad\thanks{Email: julian.konrad@tuhh.de}}
\author[b]{Janina Mittelhaus}
\author[c]{David M. Wilkins}
\author[b]{Bodo Fiedler}
\author[a,d]{Robert Meißner}

\affil[a]{Hamburg University of Technology, Institute for Interface Physics and Engineering, Am Irrgarten 3–9, 21073 Hamburg, Germany}
\affil[b]{Hamburg University of Technology, Institute of Polymers and Composites, Denickestraße 15, 21073 Hamburg, Germany}
\affil[c]{Centre for Quantum Materials and Technologies, School of Mathematics and Physics, Queen’s University Belfast, Belfast BT7 1NN, Northern Ireland, United Kingdom}
\affil[d]{Helmholtz-Zentrum Hereon, Institute of Surface Science, Max-Planck-Straße 1, 21502 Geesthacht, Germany}

\date{}

\title{Vibrational Fingerprints of Strained Polymers: A Spectroscopic Pathway to Mechanical State Prediction}

\begin{document}

\maketitle
\doublespacing

\section*{Abstract}

The vibrational response of polymer networks under load provides a sensitive probe of molecular deformation and a route to non-destructive diagnostics. Here we show that machine-learned force fields reproduce these spectroscopic fingerprints with quantum-level fidelity in realistic epoxy thermosets. Using \mbox{MACE-OFF23} molecular dynamics, we capture the experimentally observed redshifts of \textit{para}-phenylene stretching modes under tensile load, in contrast to the harmonic \mbox{OPLS-AA} model. These shifts correlate with molecular elongation and alignment, consistent with Badger’s rule, directly linking vibrational features to local stress. To capture IR intensities, we trained a symmetry-adapted dipole moment model on representative epoxy fragments, enabling validation of strain responses. Together, these approaches provide chemically accurate and computationally accessible predictions of strain-dependent vibrational spectra. Our results establish vibrational fingerprints as predictive markers of mechanical state in polymer networks, pointing to new strategies for stress mapping and structural-health diagnostics in advanced materials.



\section*{Introduction}

Non-destructive diagnostics of mechanical deformation in polymeric materials are a central challenge in the design of high-performance composites and structural systems~\cite{Mukherjee2015}. While macroscopic properties such as stiffness and toughness are routinely evaluated, molecular signatures of stress and strain remain difficult to access. Vibrational spectroscopy provides a direct pathway, revealing local structural changes through frequency shifts. Because many structural composites rely on crosslinked polymer networks, particularly epoxy resins, for strength, chemical resistance, and thermal stability~\cite{Jin2015, Balguri2021}, it is crucial to understand how mechanical load imprints on their spectroscopic features. Yet this connection remains poorly resolved.

In confined microscale domains, such as thin films between fiber layers, epoxy networks show deformation behaviors that diverge from bulk properties, including unexpected ductility under tension~\cite{Gilabert2016, Vaughan2011, Tan2018, Verschatse2020, Mittelhaus2023, Mittelhaus2025b}. These findings underscore the role of network topology and nanoscale dynamics in governing mechanical and vibrational response.

To probe these effects, molecular dynamics (MD) simulations have been widely used to investigate the atomic-scale response of crosslinked epoxy networks under load~\cite{Varshney2008, Okabe2016, Meiner2020, Konrad2021, Zhu2022, Mittelhaus2025b}. While their mechanical performance has been extensively modeled, strain-dependent vibrational behavior -- especially in confined environments~\cite{Esse2025} -- remains far less explored.

This gap is partly due to the limitations of current computational approaches for modeling vibrational spectra. Quantum chemical methods such as density functional theory yield accurate vibrational frequencies for isolated molecules~\cite{Leach2001, Jensen2016}, and have been used to characterize thermoset monomers in the gas phase~\cite{Doblies2021, Mittelhaus2023}. However, they do not capture the influence of the surrounding network. Density-functional perturbation theory extends this capability to condensed phases~\cite{Dabo2007}, but with a computational expense that confines applications to small and idealized systems.

\textit{Ab-initio} MD offers a more realistic sampling of vibrational dynamics in the condensed phase by incorporating finite-temperature and many-body effects~\cite{Praprotnik2004, Thomas2013, Zhang2023}, but it remains computationally prohibitive for large crosslinked networks. Classical MD with empirical force fields such as \mbox{OPLS-AA}~\cite{Jorgensen1996} can access larger systems, but their treatment of bonds as harmonic and their neglect of polarization and charge redistribution~\cite{Karplus2002} limit accuracy for vibrational spectra.

Machine-learned (ML) potentials bridge this gap by reproducing quantum-level accuracy at MD scales. Approaches such as Behler–Parrinello networks, Gaussian approximation potentials, and equivariant message-passing models like NequIP have demonstrated broad success~\cite{Bartk2010, Behler2016, Schran2020, Unke2021, Batzner2022, Musaelian2023}. The \mbox{MACE-OFF23} force field~\cite{Kovacs2023_ff}, developed for organic chemistry, captures anharmonicity, topological complexity, and local environmental effects, making it well-suited for polymer systems. ML potentials have recently been applied to diverse condensed-phase systems, ranging from silicates at extreme conditions~\cite{Deng2023} to thermosets~\cite{Yu2024}, expanding their role in materials science. However, prior studies used alternative ML formalisms, and \mbox{MACE-OFF23} has not yet been applied to thermosets.

Beyond ML force fields, recent advances in machine-learned models enable prediction of dipole moments and polarizabilities directly from molecular configurations~\cite{Wilkins2019, Veit2020, Kapil2020, Schienbein2023,Musaelian2023,Jana2024}. Such models extend the spectroscopic fidelity of classical and neural-network dynamics by providing IR-intensity information otherwise inaccessible in fixed-charge or non-polarizable potentials. Combining ML force fields such as MACE with symmetry-adapted Gaussian process regression (SAGPR)~\cite{Grisafi2018} enables chemically accurate simulations of both vibrational frequencies and intensities in extended polymer networks.

In this paper, we investigate the vibrational properties of crosslinked epoxy networks under uniaxial tension using MD simulations and experimental \textit{in-situ} tensile tests on film samples made from bisphenol A diglycidyl ether (DGEBA) and an aliphatic hardener to verify the simulation results. We analyze redshifts in the \textit{para}-phenylene stretching modes and bending vibrations of aromatic hydrogens under load~\cite{Mittelhaus2025a}. For the MD simulations, two epoxy systems, each composed of a different resin monomer and either an aliphatic or an aromatic hardener, are examined with both the classical \mbox{OPLS-AA} force field and the ML potential \mbox{MACE-OFF23}. In addition, we train a machine-learned dipole model on chemically representative epoxy fragments and apply it to fragment-level simulations to provide IR-intensity information complementary to velocity-based spectra. Together, these approaches bridge the gap between molecular-scale structure and vibrational response under strain in crosslinked polymer networks. By combining charge-weighted VACF analysis, ML potentials, and SAGPR-derived dipoles, we resolve strain-induced modifications to aromatic vibrational signatures. This framework reveals the molecular origins of spectral changes under stress, establishing a foundation for non-destructive diagnostics and reliability assessment in polymer networks.

\section*{Methods and Models}

To construct the molecular models, we initialized all systems from SMILES representations~\cite{Weininger1988} of the monomers bisphenol F diglycidyl ether (BFDGE), bisphenol A diglycidyl ether (DGEBA), diethyltoluenediamine (DETDA), and diethylenetriamine (DETA). The molecular structures were parameterized with the \mbox{OPLS-AA} force field, assigned via LigParGen~\cite{Dodda2017}, and we refined atomic charges using restrained electrostatic potential (RESP) fitting. We simplified molecular topologies with the TEMPLATER tool and generated reduced force fields and reaction templates following Konrad and Meißner~\cite{Konrad2025a}.

We used LAMMPS~\cite{Thompson2022} to model the crosslinking reactions and bulk polymer formation of the BFDGE-DETDA and DGEBA-DETA epoxy networks. Each simulation box contained 160 resin molecules (BFDGE or DGEBA) and 80 DETDA or 64 DETA hardener molecules, yielding total system sizes of 9360 (BFDGE-DETDA) and 9120 (DGEBA-DETA) atoms. We equilibrated the systems in the NPT ensemble for 2\,\si{ns}, followed by a crosslinking procedure performed over 2\,\si{ns} at 600\,K using the LAMMPS REACTION package~\cite{Gissinger2017}. In case of a reaction happening, we used a time step of 0.5\,\si{fs}, and applied the \texttt{NVE/limit} integrator for 500\,\si{fs} to constrain reactive-site displacements to 0.0015\,{\AA} per \si{fs}.

We applied Nosé–Hoover thermostat and barostat coupling with damping constants of 100\,\si{fs} and 1000\,\si{fs}, respectively, at a pressure of 1\,\si{atm}~\cite{Evans1985}. Periodic boundary conditions were imposed in all directions. We computed Lennard-Jones interactions with an 8\,{\AA} cutoff and short-range electrostatics with a 12\,{\AA} cutoff. Long-range electrostatics were treated using the particle–particle particle–mesh method with a (relative) $k$-space accuracy of $1.0 \times 10^{-4}$~\cite{Hockney1988}.

Following the crosslinking process, we equilibrated each cured epoxy network for 2\,ns at 300\,K to allow structural relaxation. We computed elastic moduli via linear response theory based on equilibrium trajectory sampling under a triclinic barostat over a 10\,ns interval, in accordance with prior studies~\cite{Meiner2020, Konrad2021}. To assess the mechanical response, we applied uniaxial tensile deformation up to a total strain of 0.5\,\%, using a constant strain rate of $\dot{\epsilon} = 5 \cdot 10^7\,\mathrm{s}^{-1}$. To reduce potential strain-rate artifacts, we conducted additional constant-strain relaxation simulations at intermediate strain levels of 0.03, 0.05, 0.125, 0.25, 0.375, and 0.5, each maintained for 5\,ns.

To analyze the vibrational properties, we recorded atomic velocities from equilibrated structures over 50\,ps with a 0.25\,fs time step, employing both the classical \mbox{OPLS-AA} force field~\cite{Jorgensen1996} and the ML potential \mbox{MACE-OFF23} (small) ~\cite{Batatia2022}, which captures anharmonic effects through its quantum-mechanical training data. The vibrational power spectra $S(\omega)$ were obtained from the velocity autocorrelation function (VACF) using the Wiener–Khinchin theorem. The VACF was computed as:
\begin{align}
    C_{\mathbf{v}\cdot\mathbf{v}}(\tau) &= \frac{1}{N} \sum_{i=1}^{N} \left\langle \boldsymbol{v}_i(t) \cdot \boldsymbol{v}_i(t + \tau) \right\rangle,
\end{align}
and its Fourier transform yielded the vibrational spectrum:
\begin{align}
    S(\omega) &= \int_0^{\infty} C_{\mathbf{v}\cdot\mathbf{v}}(\tau)\, e^{-i \omega \tau}\, d\tau ,
    \label{eq:vacf}
\end{align}

While IR spectra are conventionally derived from dipole autocorrelations, fixed-charge classical force fields do not account for dynamic charge redistribution. The total dipole moment for systems with fixed atomic charges $q_i$ is given by:
\begin{align}
    \boldsymbol{\mu}(t) = \sum_{i=1}^{N} q_i \mathbf{r}_i(t),
\end{align}
where $\mathbf{r}_i(t)$ is the position of atom $i$ at time $t$. While this definition is valid for the total system dipole, it does not capture time-dependent polarization or charge transfer effects at the molecular level. To qualitatively account for the influence of partial charges on vibrational dynamics, we employed a charge-weighted velocity autocorrelation function:
\begin{align}
    C_{JJ}(\tau) = \frac{1}{N} \sum_{i=1}^{N} q_i \left\langle \boldsymbol{v}_i(t) \cdot \boldsymbol{v}_i(t + \tau) \right\rangle.
    \label{eq:qvacf}
\end{align}
This approach heuristically emphasizes the motion of atoms with larger partial charges, which are typically located in polar functional groups. These atoms experience stronger local electric field fluctuations during molecular vibrations and thus tend to contribute more prominently to time-dependent dipole fluctuations. Although dynamic polarization is not explicitly modeled, $C_{JJ}(\tau)$ provides a computationally efficient and physically motivated way to improve the interpretability of vibrational spectra from fixed-charge MD simulations.

Starting from the monomeric structures of epoxy precursor moieties (Figure~\ref{fig:epoxy_structures}), we built simulation systems large enough to capture bulk polymer behavior, while remaining within the computational limits of neural network potentials and the data needs of VACF-based vibrational analysis. Each system contained about 10,000 atoms, with three velocity components per atom, sampled over 200,000 time steps (spanning 50\,ps with a 0.25\,fs interval). To achieve high wavenumber resolution, we evaluated the autocorrelation over the full lag range of 50\,ps. The resulting memory and compute demands necessitated parallel implementation on GPU architectures, and all analyses were performed on NVIDIA A100 and H100 units equipped with 80\,GB of RAM. All spectra were broadened using Gaussian convolution with a width $\sigma$ of 5\,cm$^{-1}$ to facilitate peak identification and improve comparability.

To generate a representative training dataset for machine learning–based dipole prediction, we systematically sampled chemically relevant substructures of the epoxy systems. The selected units comprised monomeric BFDGE and DETDA, as well as progressively crosslinked fragments: (i) a single BFDGE covalently linked to DETDA, (ii) two BFDGE molecules doubly connected to a single DETDA, and (iii) a BFDGE unit bridging two DETDA molecules. These fragments capture the key chemical environments of the bulk networks and provide transferable descriptors for ML dipole models.

We generated conformational ensembles of each fragment using replica exchange molecular dynamics (REMD) with the \mbox{OPLS-AA} force field. Temperature windows spanned 300–580\,K in 40\,K increments, ensuring coverage of the relevant configurational space. Each REMD trajectory ran for 500\,ns, providing sufficient sampling of both local torsional motions and large-scale rearrangements. We extracted structures for QM calculations exclusively from the 300\,K replica to retain thermally accessible conformers at ambient conditions. To reduce redundancy, we applied farthest-point sampling in the space of 17 dihedral descriptors, which reflect the dominant internal degrees of freedom of the fragments. This yielded 5000 unique structures per fragment, producing a diverse and balanced dataset.

All QM reference calculations were performed with ORCA 6.1.1~\cite{ORCA6} package at the B3LYP/def2-TZVP level of theory. For each structure, we extracted atomic charges, atomic dipole moments, and total molecular dipole vectors.

The molecular dipole moments $\boldsymbol{\mu}$ served as training data for a symmetry-adapted Gaussian process regression (SAGPR) model~\cite{Grisafi2018,Deringer2021}. SAGPR extends standard Gaussian process regression (GPR) to learn properties that transform covariantly under rigid-body rotations.
For a molecule $\mathcal{M}$, the Cartesian components of the predicted dipole moment $\boldsymbol{\mu}(\mathcal{M})$ are given by,
\begin{equation}
\mu_{i}(\mathcal{M}) = \sum_{m,j} K_{ij}(\mathcal{M},\mathcal{M}_{m}) w_{mj},
\end{equation}
where the $m$ index runs over all members of an active set as in GPR~\cite{Rasmussen2005} and $\boldsymbol{K}(\mathcal{M},\mathcal{M}')$ is a matrix-valued kernel between two molecules $\mathcal{M}$ and $\mathcal{M}'$, encoding both the similarity of these two molecules and the symmetry of the problem.
When the molecular coordinates are rotated by a given matrix $\boldsymbol{R}$, the kernel matrix is transformed as,
\begin{equation}
K_{ij}(\boldsymbol{R}\mathcal{M},\boldsymbol{R}'\mathcal{M}') = \sum_{kl}R_{ik} R_{jl}' K_{kl}(\mathcal{M},\mathcal{M}'),
\end{equation}
ensuring that the prediction transforms in the correct way under a rotation. We trained a single SAGPR model on 80\% of the molecular fragment dataset. This model, referred to as $\boldsymbol{\mu}$-EPOXY, is described in detail in the Supplementary Information (Section~S2), including the chosen hyperparameters and usage instructions.

To evaluate the performance of the trained SAGPR model in vibrational spectroscopy, we selected a DETDA–BFDGE–DETDA (DBD) fragment as a representative epoxy subunit. This configuration captures the key chemical motifs of the crosslinked networks and served as a benchmark for spectroscopic analysis under molecular strain. Molecular dynamics simulations were carried out with the \mbox{MACE-OFF23} (small) ML potential~\cite{Batatia2022}. To probe the vibrational response under finite deformation, we applied ten increments of fixed molecular elongation, systematically straining the DBD fragment. At each elongation step, trajectories were propagated for 50\,ps with a 0.25\,fs time step.

The vibrational spectra were derived via the standard autocorrelation–Fourier transform workflow. Specifically, we computed velocity autocorrelation functions (Equation~\ref{eq:qvacf}) from the velocity trajectories and Fourier transformed them to obtain vibrational power spectra. In addition, the same trajectories were used to generate instantaneous atomic dipoles from the SAGPR model. From these, we computed a dipole autocorrelation function (DACF),
\begin{align}
    C_{\boldsymbol{\mu}\cdot\boldsymbol{\mu}}(\tau) = \frac{1}{N} \sum_{i=1}^{N} \left\langle \boldsymbol{\mu}_i(t) \cdot \boldsymbol{\mu}_i(t + \tau) \right\rangle,
    \label{eq:dacf}
\end{align}
where $\boldsymbol{\mu}_i(t)$ denotes the instantaneous dipole contribution of atom $i$. Fourier transformation of the DACF, followed by multiplication with $\omega^2$, yielded the corresponding IR spectrum.

The experimental setup combined tensile testing with \textit{in-situ} infrared spectroscopy. We prepared epoxy thin films (\SI{30}{\micro\metre} thick) using an infusion process described in detail elsewhere~\cite{Mittelhaus2023,Mittelhaus2025b}. The resin system consisted of EPIKOTE™ Resin MGS RIMR 135 (diglycidyl ether of bisphenol A, DGEBA) and EPIKURE™ Curing Agent MGS RIMH 137 (aliphatic diamine hardener), mixed at a 100:30 weight ratio. We punched the films into dogbone specimens (for details on the sample geometry, see~\cite{Mittelhaus2025b}) and conducted uniaxial tensile tests at room temperature on a Deben \textit{MT200} microtensile stage with a \SI{20}{\newton} load cell, applying a crosshead speed of \SI{1.0}{\milli\metre\per\minute}. For \textit{in-situ} infrared measurements, we integrated the tensile stage into a Bruker TENSOR II spectrometer. We collected spectra in transmission mode every 7\,s during deformation and analyzed stress-sensitive vibrational modes by Gaussian peak fitting to extract peak positions and track load-induced shifts~\cite{Mittelhaus2025b}.

\section*{Results and Discussions}

\begin{figure}[htb!]
    \centering
    \includegraphics[width=0.6\textwidth]{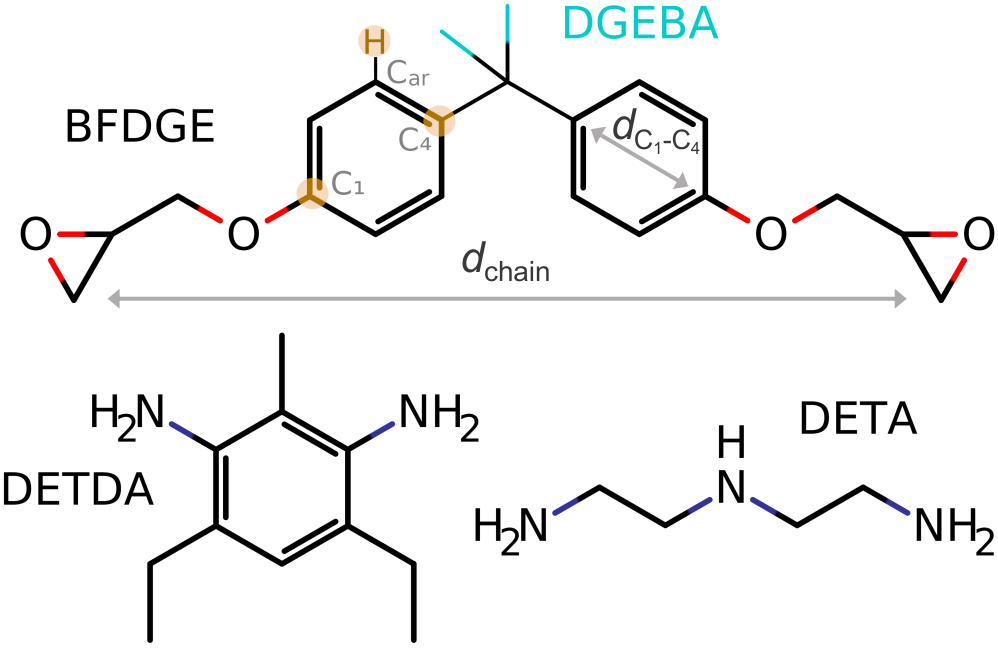}
    \caption{Chemical structures of the epoxy networks studied. The monomers BFDGE and DGEBA are shown, with additional methyl groups in DGEBA highlighted in cyan. The structures of the hardeners DETDA and DETA are also displayed. Atoms relevant to aromatic vibrational modes -- specifically the $\mathrm{C}_1$ and $\mathrm{C}_4$ positions on the aromatic rings and one $\mathrm{C}_{\mathrm{ar}} \text{--} \mathrm{H}$ group -- are highlighted in orange.}
    \label{fig:epoxy_structures}
\end{figure}

We validated the BFDGE–DETDA and DGEBA–DETA (Figure~\ref{fig:epoxy_structures}) epoxy networks for structural and mechanical properties. Their densities, elastic moduli, stress–strain behavior, and viscoelastic relaxation profiles show good agreement with experimental and prior simulation data~\cite{Tao2006, Cai2008, Littell2008, Hexion, Kallivokas2019, Meiner2020, Konrad2021, Delasoudas2024, Winetrout2024}. Full validation details are provided in the Supplementary Information (Section~S1).

From the equilibrated structures obtained after constant-strain relaxation, we sampled atomic velocities using both the classical \mbox{OPLS-AA} force field~\cite{Jorgensen1996} and the ML potential \mbox{MACE-OFF23}~\cite{Batatia2022}. Based on the resulting trajectories and RESP-fitted atomic charges, we computed the charge-weighted velocity autocorrelation function ($C_{JJ}(\tau)$; Equation~\ref{eq:qvacf}) and obtained vibrational spectra by Fourier transformation.

\begin{figure}[htb!]
    \centering
    \includegraphics[width=1.0\textwidth]{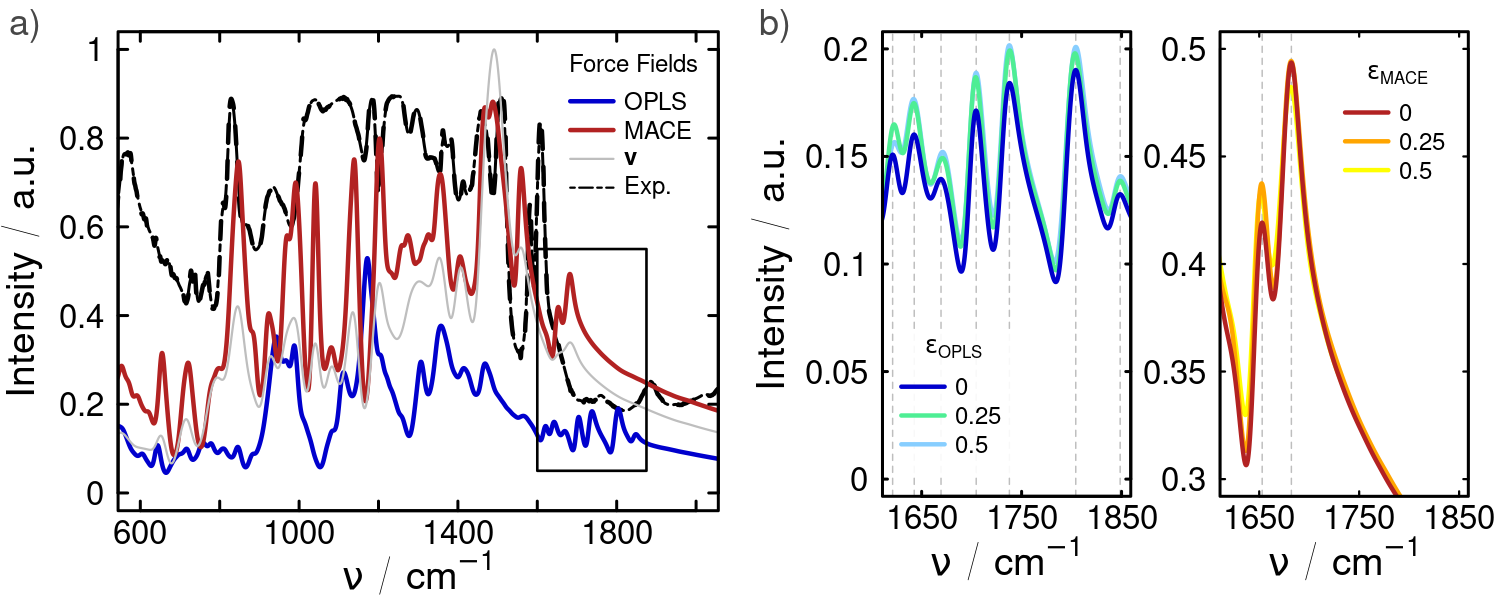}   
    \caption{Comparison of vibrational spectra derived from different force fields. a) The 600–1800\,cm$^{-1}$ region highlights differences between the OPLS-AA and MACE-OFF23 potentials. Gray curves show spectra from the classical VACF (Equation~\ref{eq:vacf}), illustrating the lower peak resolution of this approach. Colored curves show spectra from the charge-weighted VACF$_q$, with sharper, more assignable peaks. The black dashed line represents the experimental spectrum, which is slightly redshifted due to the macroscopic electrostatic environment. b) Close-up of the wavenumber region where \textit{para}-phenylene vibrations are expected.}
    \label{fig:spectrum_comparison}
\end{figure}

\begin{table}[htbp]
    \centering
    \caption{Vibrational wavenumbers derived from the charge-weighted velocity autocorrelation function (VACF$_q$) for the \mbox{OPLS-AA} and \mbox{MACE-OFF23} force fields at different strain levels ($\epsilon$) in the BFDGE-DETDA system. The indices (1--7) correspond to the peak positions identified in Figure~\ref{fig:spectrum_comparison}b.}
    \label{tab:ff_wavenumbers}
    \renewcommand{\arraystretch}{1.2}
    \begin{tabular}{c|c|c|c|c|c|c|c}
         & \multicolumn{7}{c} {$\nu_\mathrm{OPLS}$ / $\mathrm{cm}^{-1}$} \\
        \hline
        $\epsilon$ & $\nu_1$ & $\nu_2$ & $\nu_3$ & $\nu_4$ & $\nu_5$ & $\nu_6$ & $\nu_7$ \\
        \hline
        0 & \cellcolor{gray!25}1621 & \cellcolor{gray!25}1642.6 & \cellcolor{gray!25}1669.4 & \cellcolor{gray!25}1704.6 & \cellcolor{gray!25}1737.8 & \cellcolor{gray!25}1804.2 & \cellcolor{gray!25}1848.6 \\
        0.25 & \cellcolor{blue!25}1622.6 & \cellcolor{red!25}1642.2 & \cellcolor{blue!25}1671 & \cellcolor{red!25}1704.2 & \cellcolor{blue!25}1738.2 & \cellcolor{red!25}1803.4 & \cellcolor{gray!25}1848.6 \\
        0.5 & \cellcolor{blue!25}1623 & \cellcolor{red!25}1641.8 & \cellcolor{gray!25}1671 & \cellcolor{blue!25}1704.6 & \cellcolor{red!25}1737.8 & \cellcolor{blue!25}1804.2 & \cellcolor{blue!25}1849 \\
        \hline
         & \multicolumn{7}{c} {$\nu_\mathrm{MACE}$ / $\mathrm{cm}^{-1}$} \\
        \hline
        0 & \cellcolor{gray!25}1653 & \cellcolor{gray!25}1682.2 & --- & --- & --- & --- & --- \\
        0.25 & \cellcolor{red!25}1652.6 & \cellcolor{red!25}1681.8 & --- & --- & --- & --- & --- \\
        0.5 & \cellcolor{red!25}1651.8 & \cellcolor{red!25}1681.4 & --- & --- & --- & --- & --- \\
        \hline
        \end{tabular}
\end{table}

We observed significant differences in the wavenumber region associated with \textit{para}-phenylene vibrations (Figure~\ref{fig:spectrum_comparison}). The carbon atoms of the \textit{para}-phenylene unit responsible for these modes -- highlighted in Figure~\ref{fig:epoxy_structures} -- correspond to features also detected in experimental IR spectra~\cite{Mittelhaus2025b}. In the 1600–1800\,cm$^{-1}$ region, where aromatic vibrations are expected, the \mbox{OPLS-AA} model produced multiple peaks that could not be clearly assigned and showed no systematic strain dependence. This behavior reflects the harmonic spring parameters used for aromatic bonds, which generate coupled modes and diffuse spectral features. In contrast, the \mbox{MACE-OFF23}-derived spectrum showed two well-resolved peaks: one near 1650\,cm$^{-1}$, assigned to the symmetric stretch of the \textit{para}-phenylene group, and another around 1680\,cm$^{-1}$, associated with aromatic vibrations in the DETDA moiety of the polymer network. These features agree with experimental and quantum data~\cite{Mittelhaus2025b}, confirming the improved spectral fidelity of the \mbox{MACE-OFF23} model.

The strain-dependent peak positions further illustrate these differences (Table~\ref{tab:ff_wavenumbers}). As expected from its harmonic nature, \mbox{OPLS-AA} yields essentially unchanged frequencies with increasing strain. By contrast, \mbox{MACE-OFF23} predicts a clear and systematic redshift. This behavior reflects anharmonic bond softening under load, which harmonic force fields cannot reproduce. Because \mbox{MACE-OFF23} is trained on quantum-level data, it captures the curvature of the potential energy surface and thus the correct strain-induced frequency shifts. The contrast highlights the inadequacy of harmonic potentials for vibrational properties, while demonstrating that ML potentials provide significantly improved fidelity. This is particularly evident in the redshift of the $\mathrm{C}_1$–$\mathrm{C}_4$ mode of the aromatic \textit{para-}phenylene ring, which emerges as a sensitive spectral fingerprint of local strain.

To enhance the accuracy of the vibrational spectra and facilitate more precise assignments, we isolated the velocities of the $\mathrm{C}_1 \text{--} \mathrm{C}_4$ and $\mathrm{C}_{\mathrm{ar}} \text{--} \mathrm{H}$ atoms. Including all atomic velocities introduces motion from atoms bonded to the aromatic ring -- such as bridging oxygens and adjacent carbons -- that couple with ring vibrations and obscure intrinsic IR modes. These contributions primarily broaden and shift peaks without adding information.

\begin{figure}[htb!]
    \centering
    \includegraphics[width=0.6\textwidth]{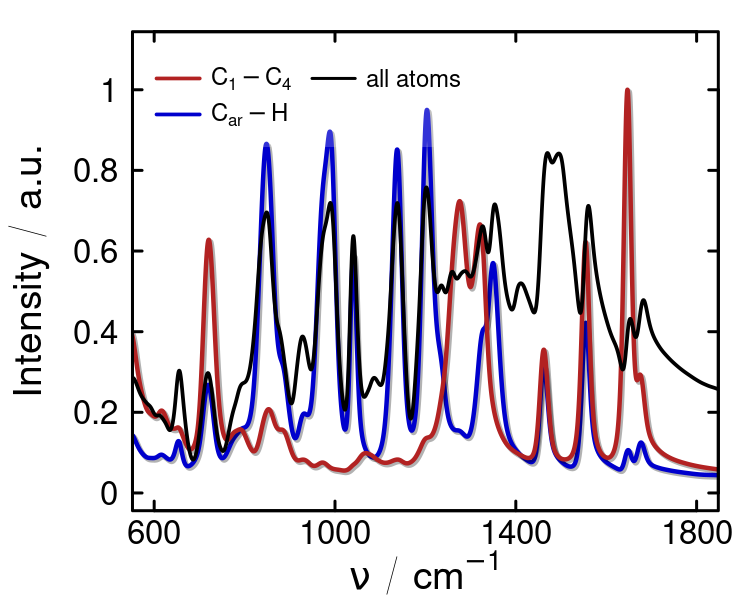}
    \caption{Vibrational spectrum showing the contributions of specific atomic groups to the total signal. The spectrum computed from all atomic velocities is shown in black. The red curve corresponds to the isolated $\mathrm{C}_1 \text{--} \mathrm{C}_4$ atoms of the aromatic ring, while the blue curve represents the hydrogen atoms attached to the \textit{para}-phenylene group.}
    \label{fig:spectrum_resid}
\end{figure}

By isolating the aromatic contributions through evaluation of the charge-weighted velocity autocorrelation function using only the velocities of the aromatic atoms (Figure~\ref{fig:spectrum_resid}), we obtained spectra that selectively emphasize aromatic vibrational modes. The analysis focused on characteristic features: the $\mathrm{C}_1 \text{--} \mathrm{C}_4$ stretching modes (shown in red), comprising a symmetric stretch near 1650\,cm$^{-1}$ and an asymmetric stretch around 1550\,cm$^{-1}$, and the $\mathrm{C}_{\mathrm{ar}} \text{--} \mathrm{H}$ bending modes (shown in blue), including an out-of-plane (oop) vibration at approximately 850\,cm$^{-1}$ and an in-plane (ip) vibration near 990\,cm$^{-1}$~\cite{Andrejeva2016}. This decomposition sharpened peak assignment and clarified the molecular origins of the aromatic features within the epoxy network.

\begin{figure}[htb!]
    \centering
    \includegraphics[width=1.0\textwidth]{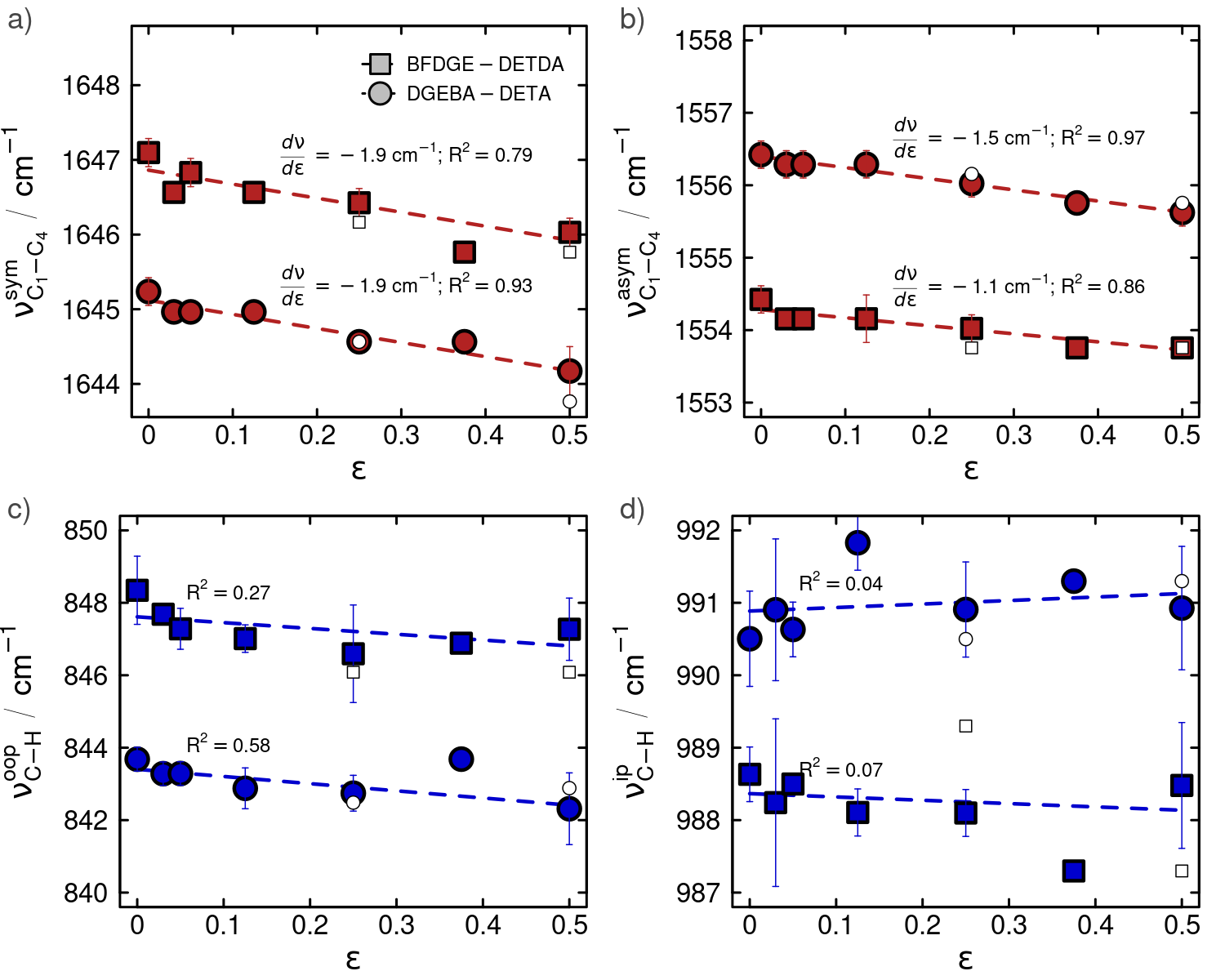}
    \caption{Vibrational peak shifts of the \textit{para}-phenylene group under increasing strain for both epoxy systems, with linear fits applied to the extracted data. Error bars indicate the standard deviation across deformation along the $x$-, $y$-, and $z$-directions. White data points represent simulations conducted at a tenfold reduced strain rate. a) Symmetric stretch vibration: a linear redshift was observed in both systems, closely matching the experimentally reported value of \SI{-1.5}{\per\centi\meter}~\cite{Mittelhaus2025b}, as reflected in the fitted slopes. b) Asymmetric stretch vibration: a similar redshift trend with increasing strain was observed, reinforcing the reproducibility of strain-dependent vibrational behavior in the simulations. c)/d) For both bending modes, no significant trend toward red- or blueshifting was detected across strain levels.}
    \label{fig:shift_cc}
\end{figure}

Using this method, we extracted vibrational peaks for both epoxy systems, sampled with the \mbox{MACE-OFF23} potential, under uniaxial deformation in all three spatial directions ($x$, $y$, $z$) (Figure~\ref{fig:shift_cc}). For each strain level ($\epsilon = 0$, 0.03, 0.05, 0.125, 0.25, 0.375, 0.5), spectra were computed independently for deformation along each axis, and the resulting peak positions were averaged. We included the standard deviation across directions as an error bar, capturing the mechanical anisotropy of the networks in their vibrational response.

As shown in Figure~\ref{fig:shift_cc} a) and b), we observed a consistent redshift in the \textit{para}-phenylene stretch vibrations of both epoxy systems with increasing strain. The magnitude of this shift agrees with creep experiments, which report $\frac{{\rm d}\nu_{\mathrm{exp}}}{{\rm d}\epsilon} = -1.5,\mathrm{cm}^{-1}$~\cite{Mittelhaus2025b}. At zero strain, the BFDGE–DETDA system showed a slightly higher frequency for the symmetric stretch vibration compared to DGEBA–DETA, whereas the opposite trend was observed for the asymmetric stretch, where DGEBA–DETA exhibited a higher-frequency peak. This difference likely arises from the methyl substituents on the DGEBA aromatic rings (see Figure~\ref{fig:epoxy_structures}, highlighted in cyan), which modify the local environment of the symmetric and asymmetric modes in opposite ways. Despite these differences in absolute wavenumbers, the overall redshift behavior under strain was consistent across both systems, reinforcing the predictive reliability of the simulations.

\begin{figure}[htb!]
    \centering
    \includegraphics[width=1.0\textwidth]{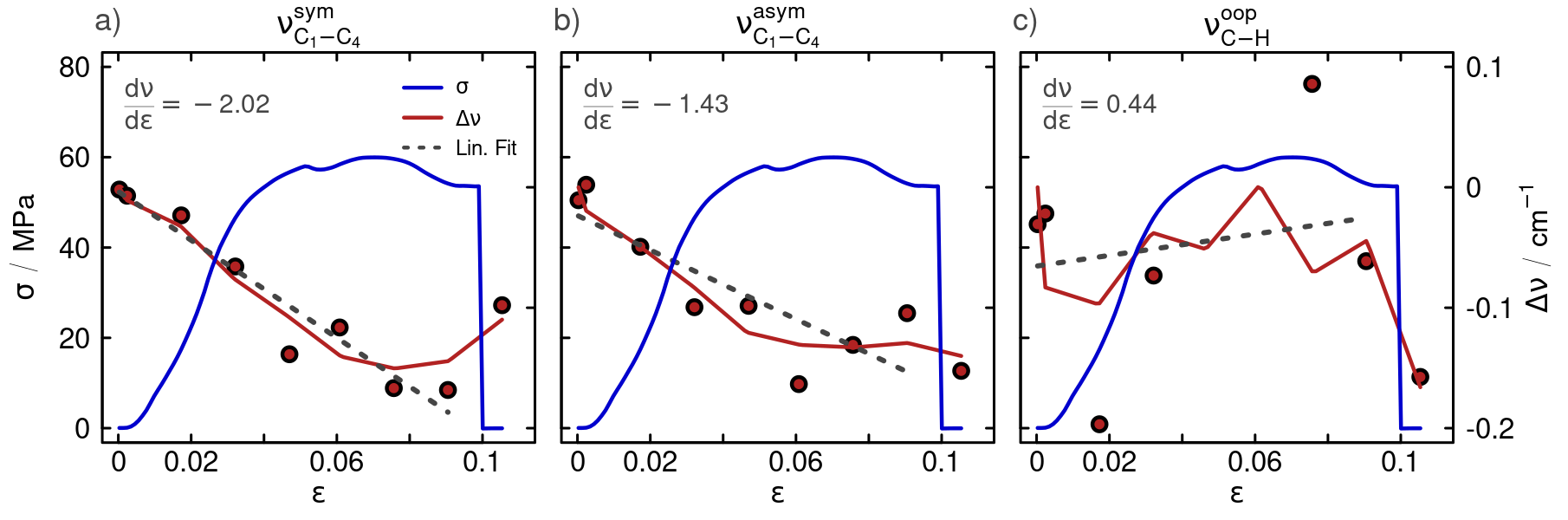}
    \caption{Experimentally measured strain-dependent peak shifts obtained from \textit{in-situ} IR tensile testing of epoxy films. a) $\nu_\mathrm{C_1-C_4}^{\mathrm{sym}}$ stretch, b) $\nu_\mathrm{C_1-C_4}^{\mathrm{asym}}$ stretch, and c) $\nu_\mathrm{C-H}^{\mathrm{oop}}$ bending vibration. Blue curves represent the stress–strain response ($\sigma$), red curves the peak shift ($\Delta \nu$), and black circles the fitted peak positions. Dashed gray lines indicate linear fits with corresponding slopes ($\mathrm{d}\nu/\mathrm{d}\epsilon$).}
    \label{fig:exp_shift}
\end{figure}

In order to directly validate the simulated strain-dependent spectral shifts, we performed \textit{in-situ} infrared measurements during uniaxial tensile loading on \SI{30}{\micro\metre} epoxy films. Six independent tensile tests were carried out at different strain rates ($\dot{\epsilon}=0.2$–\SI{1.0}{\milli\metre\per\minute}). The mechanical response showed an average ultimate tensile strength of $58.6\pm5.2$~MPa and a elongation at break of $0.073\pm0.017$. The peak positions at zero strain were located at \SI{1608.3}{\per\centi\metre} ($\nu_\mathrm{C_1-C_4}^{\mathrm{sym}}$), \SI{1581.2}{\per\centi\metre} ($\nu_\mathrm{C_1-C_4}^{\mathrm{asym}}$), and \SI{830.9}{\per\centi\metre} ($\nu_\mathrm{C-H}^{\mathrm{oop}}$). Linear regression of the peak shifts yielded average slopes of $-2.56\pm0.49$~cm$^{-1}$ for the symmetric stretch, $-2.20\pm0.58$~cm$^{-1}$ for the asymmetric stretch, and $-1.30\pm1.53$~cm$^{-1}$ for the out-of-plane bending vibration. The latter mode shows only weak and inconsistent shifts. In the representative experiment shown in Figure~\ref{fig:exp_shift}, a small apparent blueshift is visible (slope $+0.44$~cm$^{-1}$), but this effect was absent in the majority of tests, confirming that bending vibrations of aromatic hydrogen atoms are far less sensitive to tensile deformation than the \textit{para}-phenylene stretches in the backbones. Additionally, the corresponding redshifts of the symmetric and asymmetric stretching vibrations are in excellent agreement with the simulation data shown in Figure~\ref{fig:shift_cc}.

The mechanical properties, including ultimate tensile strength and elongation at break, showed no systematic dependence on strain rate. In contrast, the $\nu_\mathrm{C_1-C_4}^{\mathrm{sym}}$ and $\nu_\mathrm{C_1-C_4}^{\mathrm{asym}}$ stretching vibrations exhibited a tendency toward larger redshift slopes at slower strain rates, whereas the $\nu_\mathrm{C-H}^{\mathrm{oop}}$ bending vibration showed no reproducible trend. The variability across experiments attributed to sample-to-sample differences arising from the manual preparation of thin films, such as slight variations in thickness or the presence of micro-defects. The data shown in Figure~\ref{fig:exp_shift} represent a typical test at $\dot{\epsilon}=0.5$~mm/min, while additional results at different strain rates and the corresponding strain rate dependency are provided in Supplementary Section~S3.

To extend the vibrational analysis, we investigated the bending modes of the hydrogen atoms attached to the aromatic ring, as shown in Figure~\ref{fig:shift_cc} c) and d). According to the findings of Mittelhaus \textit{et al.}~\cite{Mittelhaus2025b}, stretching of the \textit{para}-phenylene group leads to a blueshift in these bending vibrations. In the present study, we analyzed the bending modes analogously to the stretches, but the peak positions were highly scattered, remaining nearly constant or showing only slight shifts to lower wavenumbers. Because the coefficients of determination (R$^2$) were low, we could not confirm a consistent blueshift. This discrepancy may be attributed to limitations of the \mbox{MACE-OFF23} potential. In particular, polarization changes associated with aromatic deformation were likely absent from the training data. Moreover, the employed ML architecture lacks explicit atomic charges, which may prevent it from capturing subtle electronic effects such as those governing shifts in polar modes.

\begin{figure}[htb!]
    \centering
    \includegraphics[width=1.0\textwidth]{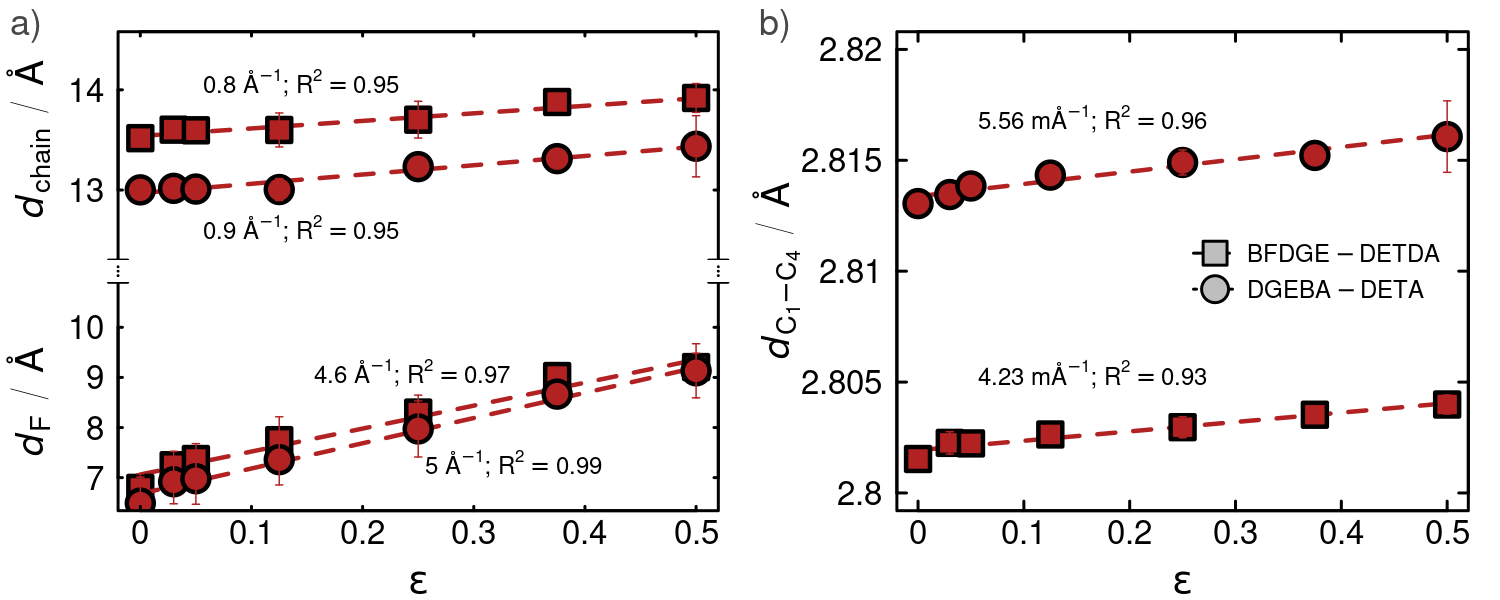}
    \caption{a) Evolution of the inter-chain distance $d_{\mathrm{chain}}$ and the aligned component of the distance vector in the direction of tension, $d_{\mathrm{F}}$, as a function of strain for both epoxy systems. b) Corresponding changes in the aromatic ring spacing $d_{\mathrm{C}_1\text{--}\mathrm{C}_4}$ with strain. The slope values indicate the rate of change with applied strain.}
    \label{fig:network}
\end{figure}

To elucidate the origin of the redshift observed in the stretching modes, we analyzed the molecular configuration of the resin molecules -- specifically their orientation and extension -- following the methodology outlined in a previous study~\cite{Mittelhaus2025b}. The elongation of the polymer backbone was quantified by the inter-terminal carbon distance, $d_{\mathrm{chain}}$, for each resin molecule. The component of this distance aligned with the direction of applied force and strain, denoted as $d_{\mathrm{F}}$, was also evaluated. As shown in Figure~\ref{fig:network} a), both $d_{\mathrm{chain}}$ and $d_{\mathrm{F}}$ increased approximately linearly with strain, indicating that the polymer network accommodated deformation through a combination of segmental extension and alignment~\cite{Mittelhaus2025b}. The quantities $d_{\mathrm{chain}}$ and $d_{\mathrm{C}_1\text{--}\mathrm{C}_4}$, used to characterize these structural changes, are schematically marked in Figure~\ref{fig:epoxy_structures}. Here, $d_{\mathrm{F}}$ denotes the component of the chain vector aligned with the applied strain direction (e.g. the $x$-axis for $\epsilon_x$): $d_{\text{chain}} = \left\lVert \begin{pmatrix} d_{\mathrm{F}} & y & z \end{pmatrix} \right\rVert$

Stronger chain alignment was observed in the DGEBA–DETA system than in BFDGE\-DETDA, as reflected by the steeper slope of $d_{\mathrm{chain}}$ and $d_{\mathrm{F}}$ in response to strain. This structural adaptation induced elongation of the \textit{para}-phenylene group, quantified by the distance between $\mathrm{C}_1$ and $\mathrm{C}_4$ atoms, $d_{\mathrm{C}_1\text{--}\mathrm{C}_4}$, which correlates with the redshift of its vibrational modes in accordance with Badger’s rule~\cite{Badger1934}. The linear relationship observed in Figure~\ref{fig:network} b) confirms that strain-induced molecular elongation of the aromatic ring directly contributes to the frequency shifts in the vibrational spectra. This trend highlights the increased stretching of \textit{para}-phenylene units in the DGEBA–DETA system, consistent with its stronger overall alignment behavior.

Although both epoxy systems exhibited broadly similar trends, subtle but meaningful differences emerged. In DGEBA–DETA, chain alignment was approximately 20\,\% more pronounced than in BFDGE–DETDA, as indicated by the slope of $d_{\mathrm{F}}$. This stronger alignment was accompanied by an increase of $d_{\mathrm{C}_1\text{--}\mathrm{C}_4}$. However, the corresponding redshift of the \textit{para}-phenylene stretching modes did not scale proportionally. This discrepancy suggests that the spectral response is governed not only by geometric elongation but also by local network flexibility, anharmonic coupling, and vibrational mode mixing.

\begin{figure}[htb!]
    \centering
    \includegraphics[width=0.6\textwidth]{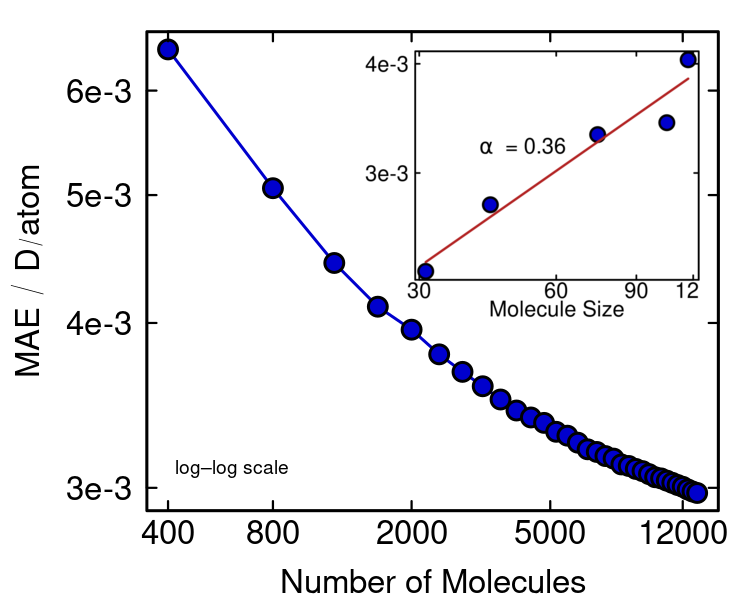}
    \caption{Learning curve of the SAGPR dipole model. The inset shows that the per-atom error increases super-linearly with molecule size, following $\varepsilon/n \sim n^{\alpha}$ with $\alpha \approx 1/3$.}
    \label{fig:learning}
\end{figure}

Figure~\ref{fig:learning} shows the learning curve for the $\boldsymbol{\mu}$-EPOXY model for molecular dipole moments. Using 80\,\% of the training set, an SAGPR model was built that gives an error of $0.00297323~\text{D/atom}$ on an independent validation set, or $\sim 5\,\%$ of the intrinsic deviation in the training set. This error is somewhat larger than that reported for recent models of water polarization ($\sim 1\,\%$)~\cite{Kapil2020,Jana2024}, reflecting the greater structural complexity and size dispersity of the epoxy fragments studied here. In the inset of Figure~\ref{fig:learning} we show the prediction error as a function of molecular size in the validation set.
The error in predicting dipole moment per atom shows a weak size dependence, increasing approximately as $\varepsilon / n \sim n^{\alpha}$ with $\alpha \approx 1/3$, where $n$ is the number of atoms and $\varepsilon$ the prediction error. This trend suggests that care should be taken when applying the model to arbitrarily large systems. Its most meaningful test of extrapolative ability, however, lies in the present application to epoxy networks.

\begin{figure}[htb!]
    \centering
    \includegraphics[width=0.6\textwidth]{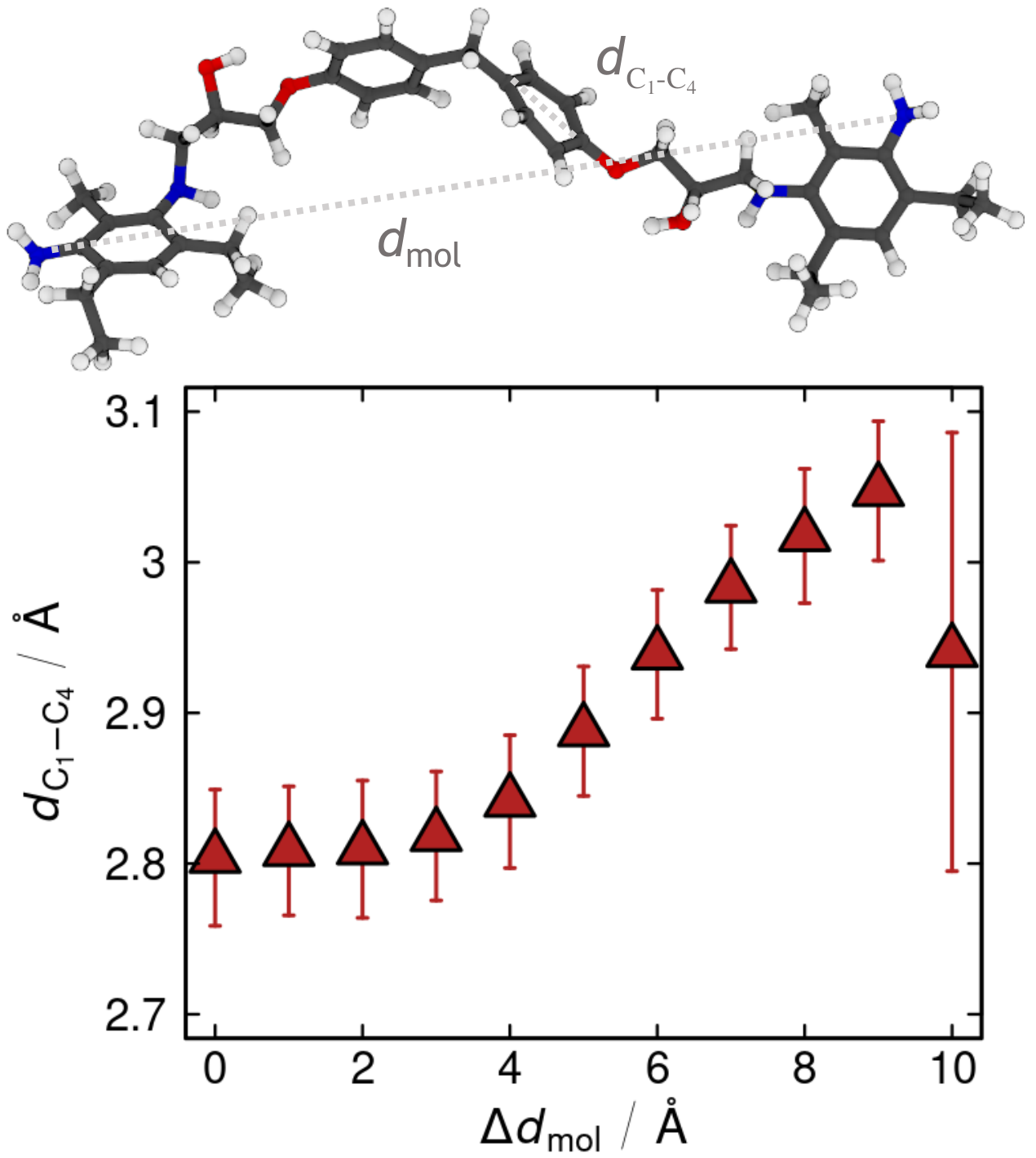}
    \caption{DBD fragment under controlled elongation. The aromatic span $d_{\mathrm{C}1\text{--}\mathrm{C}4}$ increases with $\Delta d_\mathrm{mol}$, then drops at $\Delta d_\mathrm{mol}=10\,${\AA} due to epoxy bond dissociation. Error bars indicate variability across configurations at each elongation.}
    \label{fig:dbd_dist}
\end{figure}

To connect the bulk-epoxy network analysis with the molecular-level SAGPR study, we quantified the modification of key internal distances in the DBD fragment under controlled elongation. As shown in Figure~\ref{fig:dbd_dist}, the molecular extension $\Delta d_\mathrm{mol}$ was accompanied by a concomitant increase of the aromatic ring span $d_{\mathrm{C}_1\text{--}\mathrm{C}4}$, corresponding to the same geometric lever identified in the networks (Figure~\ref{fig:network}b). At the largest imposed elongation ($\Delta d_\mathrm{mol}=10$\,{\AA}), we observed a drop in inter-fragment distance, reflecting dissociation of the initially formed epoxy linkage. This behavior indicates that substantial tensile load is borne by the \textit{para}-phenylene unit prior to bond rupture and shows that the \mbox{MACE-OFF23} potential captures bond breaking on the underlying quantum-mechanical surface. In contrast, fixed-topology classical force fields with harmonic bond terms cannot represent dissociation~\cite{Leach2001}.

\begin{figure}[htb!]
    \centering
    \includegraphics[width=0.6\textwidth]{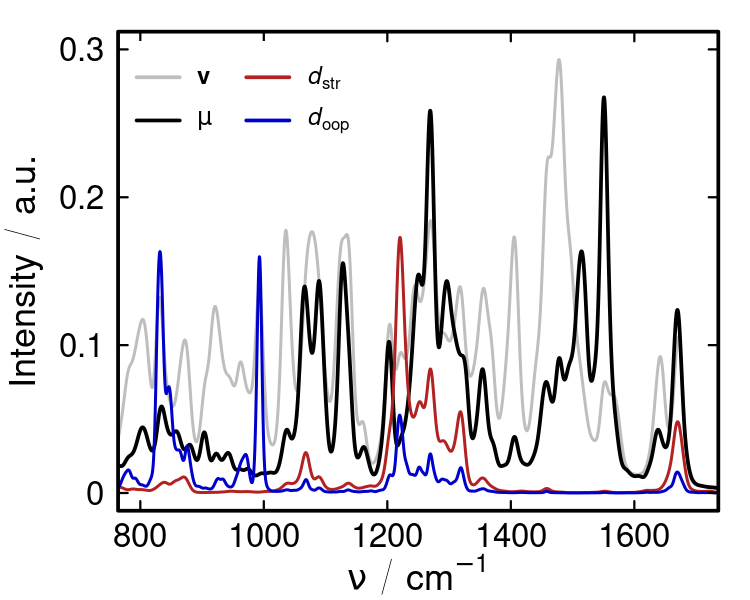}
    \caption{Comparison of vibrational spectra of the DBD fragment obtained from different autocorrelation functions. The velocity-based spectrum (gray) exhibits many features not IR-active. In contrast, the dipole-based spectrum (black) highlights only IR-active modes. Targeted ACF–FFT workflows were applied to specific internal coordinates: $\mathrm{C}_1\text{--}\mathrm{C}_4$ distance (red) for the symmetric aromatic stretch (str), and the out-of-plane (oop) motion of aromatic hydrogens (blue) for bending vibrations.}
    \label{fig:compare_spectra}
\end{figure}

The vibrational response of the DBD fragment was first analyzed by comparing power spectra derived from atomic velocities with those obtained from $\boldsymbol{\mu}$-EPOXY-predicted dipoles. As shown in Figure~\ref{fig:compare_spectra}, the velocity-based spectrum (gray) exhibits a large number of vibrational features, many of which are not infrared-active. In contrast, the dipole-based spectrum (black) highlights only the IR-active modes, thereby providing a more direct comparison with experiments. 

To assign specific peaks within the spectra, we applied targeted autocorrelation function–fast Fourier transform (ACF–FFT). For the symmetric stretching vibration of the \textit{para}-phenylene group, we sampled the $\mathrm{C}_1$–$\mathrm{C}_4$ distance and computed its power spectrum (red curve in Fig.~\ref{fig:compare_spectra}). Analogously, we isolated the out-of-plane bending motion of the aromatic hydrogens by calculating fluctuations of the corresponding distance coordinate and transforming them into the frequency domain (blue curve in Fig.~\ref{fig:compare_spectra}). This approach allowed us to unambiguously assign peaks to distinct vibrational modes, providing a direct structural interpretation of the spectral features.

\begin{figure}[htb!]
    \centering
    \includegraphics[width=0.6\textwidth]{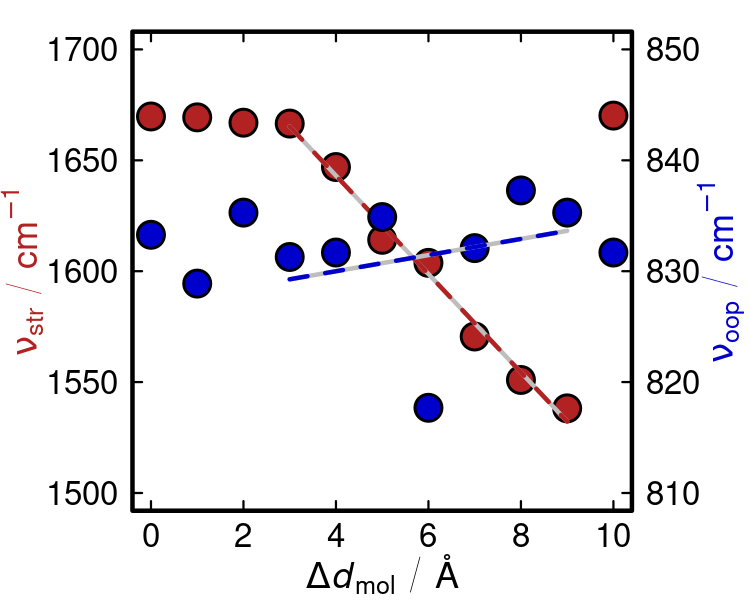}
    \caption{Tracked peak positions of the symmetric stretch vibration of the \textit{para}-phenylene group (red, left axis) and the out-of-plane bending vibration of aromatic hydrogens (blue, right axis) as a function of molecular elongation $\Delta d_\mathrm{mol}$. Linear fits highlight the redshift of the stretching mode with increasing $\Delta d_\mathrm{mol}$ and the weak blueshift of the bending mode.}
    \label{fig:compare_peaks}
\end{figure}

Peak positions were subsequently extracted from the spectra at each imposed molecular elongation. As shown in Figure~\ref{fig:compare_peaks}, the symmetric stretch of the \textit{para}-phenylene group initially appeared near 1670\,cm$^{-1}$ (red), while the out-of-plane bending vibration of the aromatic hydrogens was located around 830\,cm$^{-1}$ (blue). Tracking these frequencies as a function of $\Delta d_\mathrm{mol}$ revealed a pronounced redshift of the stretching mode, in agreement with the bulk-network analysis of the power spectra. The fitted trend line confirms that the redshift coincides with the increase of the $d_{\mathrm{C}_1\text{--}\mathrm{C}_4}$ distance, thereby establishing a direct link between molecular elongation and redshifting of the aromatic stretch vibration $\nu_{\mathrm{C}_1\text{--}\mathrm{C}_4}$.

Importantly, the out-of-plane bending vibration was also captured in the dipole-derived spectra, but not in the velocity-based spectra, underscoring the advantage of dipole autocorrelation functions for detecting IR-active modes. The identification of both stretching and bending responses shows that the dipole-enhanced workflow captures a broader range of vibrational behavior under strain, strengthening its predictive power.

The results show that the redshift of the symmetric stretch is substantially stronger than the weak blueshift of the out-of-plane bending mode. This disparity is explained by the fact that the redshift directly reflects the elongation of the $\mathrm{C}_1\text{--}\mathrm{C}_4$ distance \cite{Badger1934}, whereas the blueshift is only a secondary consequence of aromatic ring deformation, as discussed by Mittelhaus \textit{et al.}~\cite{Mittelhaus2025b}.

An additional point of significance is the demonstrated generalization capability of the $\boldsymbol{\mu}$-EPOXY model. Although the training data did not include structures with $\mathrm{C}_1\text{--}\mathrm{C}_4$ distances exceeding 3.0\,\AA, the model nevertheless reproduced the expected frequency trend under elongation with high fidelity. Thus, the model not only interpolates within its training domain but also extrapolates in a physically meaningful way.

Crucially, the consistent redshift observed in both epoxy networks (BFDGE–DETDA and DGEBA–DETA) and in the strained DBD fragment demonstrates that the \textit{para}-phenylene stretches provide robust spectral markers of mechanical strain. The agreement between MACE-based bulk spectra and SAGPR–derived fragment spectra validates the fidelity of the ML descriptions and indicates that vibrational observables can be exploited to localize stress in aromatic epoxy networks. Due to their sensitivity to molecular alignment and $\mathrm{C}_1\text{--}\mathrm{C}_4$ elongation, these modes are well suited for non-destructive strain mapping in polymer composites, providing a foundation for computationally guided diagnostics and structural-health monitoring, contingent on appropriate peak–strain calibration and accounting for network anisotropy (cf. Figure~\ref{fig:shift_cc}).

\section*{Conclusions}

We developed a computational framework to predict strain-dependent vibrational responses in crosslinked epoxy networks, providing spectroscopic access to local mechanical deformation. By comparing classical (\mbox{OPLS-AA}) and machine-learned (\mbox{MACE-OFF23}) force fields, we showed that only the ML model captures the anharmonic bond softening responsible for redshifts in aromatic stretching modes under load. These shifts, observed across two chemically distinct epoxy systems, are in qualitative agreement with earlier \textit{in-situ} IR measurements performed under different loading conditions~\cite{Mittelhaus2025b} and in quantitative agreement with the \textit{in-situ} tensile data presented here, and they correlate with molecular-scale elongation and orientation, in accordance with Badger’s rule. The demonstrated agreement between simulation and experiment supports the use of vibrational fingerprints as reliable, predictive markers of mechanical state in polymer networks.

Through atom-specific spectral decomposition and the use of charge-weighted velocity autocorrelations, vibrational signatures were linked to backbone alignment and \textit{para}-phenylene ring extension. These features serve as molecular-level indicators of mechanical strain and can be accessed through infrared spectroscopy. Although polarization effects were not fully resolved -- leading to limited accuracy in bending modes -- the MACE model provided chemically realistic and spectrally consistent results, highlighting the ability of ML force fields to capture deformation-sensitive vibrational behavior in polymers.

Complementary dipole modeling further demonstrated that IR-active features, including the out-of-plane bending of aromatic hydrogens, can be resolved with higher fidelity than velocity-based spectra alone. The agreement between bulk-network spectra (from MACE) and fragment-level dipole spectra confirms that combining force-field and dipole learning approaches provides a consistent description across scales. The $\boldsymbol{\mu}$-EPOXY model also reproduced strain-induced shifts even outside its training domain, demonstrating its potential to predict vibrational observables under deformation and its meaningful extrapolation capability.

These insights directly extend to composite applications, where epoxy matrices govern reliability under load. Linking vibrational fingerprints to mechanical strain provides a pathway for non-destructive diagnostics and lifetime assessment in structural materials.

This work represents the first demonstration of \mbox{MACE-OFF23} for epoxy networks, establishing its suitability for complex crosslinked systems beyond the small molecules and condensed phases for which it was originally trained. Our results show that ML potentials offer chemically realistic and computationally tractable access to vibrational fingerprints of mechanical state in thermoset polymers. Coupling vibrational spectroscopy with ML force fields thus provides a viable route to non-destructive diagnostics, stress mapping, and failure prediction in structural polymers.

Future extensions to other thermoset systems, supported by extensive quantum-mechanical training data, are expected to enable unified architectures that simultaneously predict interatomic forces and electronic observables. Such polarization-aware neural network models~\cite{Musaelian2023} would deliver an all-in-one description of mechanical and spectroscopic responses under deformation, enhancing transferability and predictive power.

\section*{Acknowledgments}

This research received funding from Grant No. 525597740 provided by the Deutsche Forschungsgemeinschaft (DFG, German Research Foundation).

\section*{Competing interests}

The authors have no competing interests to declare.

\section*{Author contributions}

Conceptualization: JK, RM
Methodology: JK, DW, JM, BF, RM
Investigation: JK, DW, JM
Visualization: JK
Supervision: RM, BF
Writing-original draft: JK
Writing-review \& editing: JK, DW, JM, BF, RM

\section*{Data availability}

The data supporting this study are available from the corresponding author upon reasonable request; due to their multi-terabyte size, data sharing will be coordinated accordingly.

\section*{Supplementary Materials}

The supplementary information includes: Supplementary Text, Figures~S1--S5, Tables~S1 and S2.

\bibliographystyle{ieeetr} 
\bibliography{ref.bib}

\end{document}


\linenumbers
\doublespacing
\maketitle

\section*{Supplementary Information}

\setcounter{section}{0}
\setcounter{subsection}{0}

\section{Validation of Epoxy Resin Models}
\label{sec:epoxy_models}

The epoxy networks used in this study -- BFDGE–DETDA and DGEBA-DETA -- were validated to ensure consistency with known thermoset behavior. Structural, mechanical, and viscoelastic properties were examined to confirm that the simulated systems reliably capture experimentally observed trends.

Nano-scale polymer networks were successfully generated, and their structural and mechanical properties were found to be consistent with established bulk epoxy behavior. The key computed values are summarized in Table~\ref{tab:properties} adn show excellent agreement with experimental measurements and previous simulation studies \cite{Tao2006, Cai2008, Littell2008, Hexion, Kallivokas2019, Meiner2020, Konrad2021, Winetrout2024}. Bulk properties were extracted directly from MD simulation trajectories, and elastic moduli were derived using linear response theory, as previously established \cite{Konrad2021, Konrad2021prop}.

\begin{table}[tb]
    \centering
    \caption{Structural and mechanical properties of the epoxy networks.}
    \label{tab:properties}
    \renewcommand{\arraystretch}{1.2}
    \begin{tabular}{|c||c||c|}
        \hline
        & \multicolumn{1}{c||} {BFDGE-DETDA} & \multicolumn{1}{c|} {DGEBA-DETA} \\
        \hline
        $\eta$ / \% & 96 & 94  \\
        $\rho$ / $\mathrm{g} \cdot \mathrm{cm}^{-3}$ & 1.05 & 1.04  \\
        $l_{\mathrm{Box}}$ / \si{nm} & 4.6 & 4.6 \\
        \hline
        $Y$  / GPa & \num{2.4} $\pm$ \num{0.1}  & \num{2.7} $\pm$ \num{0.3}  \\
        $G$  / GPa & \num{0.9} $\pm$ \num{0.05}  & \num{1.0} $\pm$ \num{0.1}  \\
        $B$  / GPa & \num{3.1} $\pm$ \num{0.4}  & \num{3.4} $\pm$ \num{0.3}  \\
        \hline
    \end{tabular}
\end{table}

\begin{figure}[tb]
    \centering
    \includegraphics[width=1.0\textwidth]{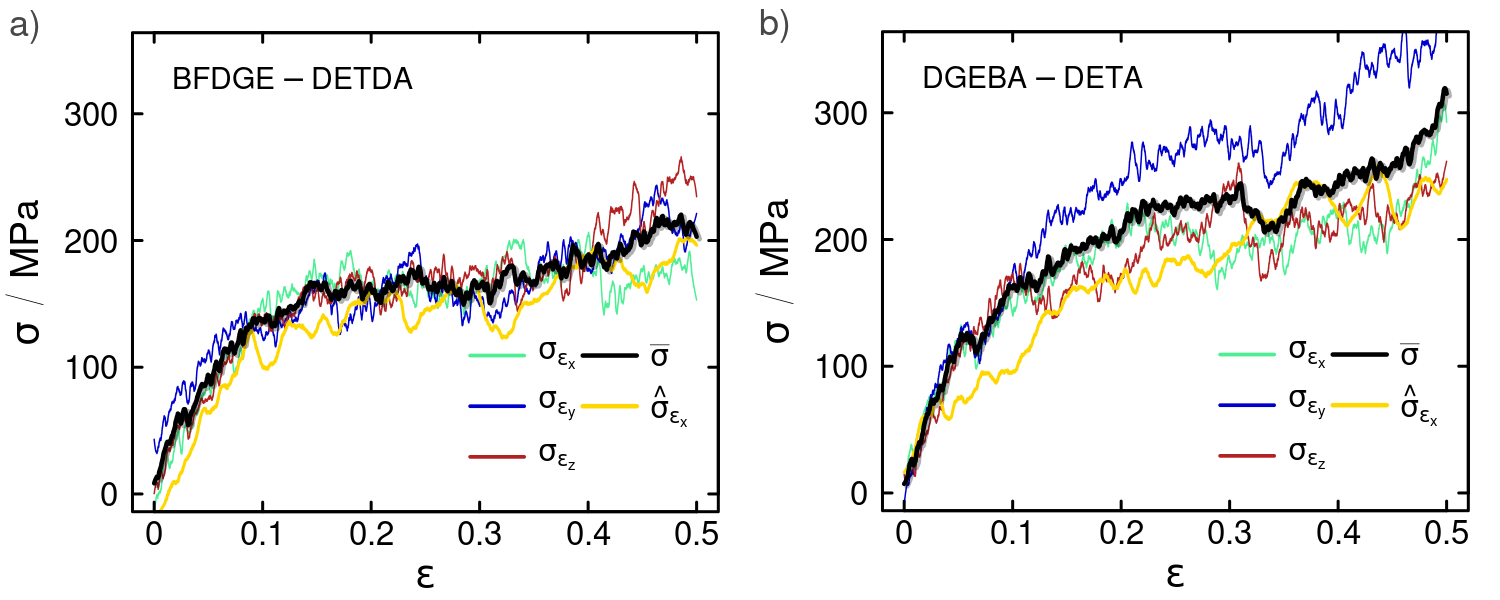}
    \caption{Stress–strain responses of the BFDGE-DETDA and DGEBA-DETA epoxy systems under uniaxial tensile deformation along the $x$-, $y$-, and $z$-directions. Curves are color-coded by direction (green, blue, red), with additional data shown in yellow representing deformation at a tenfold slower strain rate along the $x$-axis. The DGEBA-DETA system exhibits greater anisotropy compared to BFDGE-DETDA, with higher mean stress values, particularly beyond the non-linear elastic regime, reflecting the differences in network architecture.}
    \label{fig:stress_strain_comparison}
\end{figure}

The stress–strain behavior of the epoxy networks under uniaxial tensile deformation is shown in Figure~\ref{fig:stress_strain_comparison}. Simulations were carried out along all three Cartesian directions ($x$, $y$, and $z$) to assess directional mechanical response. To evaluate potential strain rate effects, an additional simulation was performed along the $x$-axis using a tenfold lower strain rate of $\dot{\epsilon} = 5 \cdot 10^6\,\mathrm{s}^{-1}$, with results indicated in yellow.

A nearly isotropic mechanical response was observed for the BFDGE-DETDA network, with stress–strain curves along all directions closely following the ensemble average. In contrast, the DGEBA-DETA system exhibited marked anisotropy, particularly at higher strain levels ($\epsilon > 0.1$), where directional differences in stress response became increasingly pronounced.

These variations in mechanical anisotropy were attributed to differences in network topology. The DGEBA-DETA system consistently sustained higher stress under load than BFDGE-DETDA, especially beyond the non-linear elastic regime. This behavior was linked to the higher cross-linking density of DGEBA-DETA, in which each DETA molecule typically connects five resin units, forming a more rigid and spatially constrained network. Conversely, the BFDGE-DETDA network, formed through four-way connections via DETDA, allowed for greater molecular mobility through chain untangling and uncoiling. This increased flexibility facilitated strain accommodation, resulting in a comparatively softer mechanical response.

To assess time-dependent viscoelastic behavior, constant-strain relaxation simulations were performed at selected strain levels up to $\epsilon = 0.5$. Each network was held at fixed strain and allowed to relax under constant volume and temperature conditions. The resulting stress decay curves were fit using a first-order exponential model~\cite{Konrad2021, Konrad2021prop}:
\begin{align}
\sigma(t) = \sigma_0 \cdot \exp\left(-\frac{t}{\tau}\right) + \sigma_{t\rightarrow\infty}.
\end{align}
The extrapolated long-time stresses, $\sigma_{t\rightarrow\infty}$, are reported in Table~\ref{tab:fit_sigma}. Values between 100\,MPa and 150\,MPa were obtained, significantly lower than the instantaneous stresses observed during dynamic loading. Although these values remain above typical experimental tensile strengths (70–80\,MPa) \cite{Littell2008, Mittelhaus2023, Mittelhaus2025b}, the discrepancy is compartively small and is attributed to the relatively small system sizes -- reducing the probability of finding structural defects isotropically -- and the limitations of the force fields used, which do not account for bond dissociation or fracture events. These effects are expected to become increasingly relevant at elevated strain levels approaching $\epsilon = 0.5$.

\begin{table}[tb]
    \centering
    \caption{First-order kinetic fits of the relaxed stresses $\sigma_{t\rightarrow\infty}$ obtained from constant strain relaxation simulations at various strain levels $\epsilon$ for each spatial direction. Values in parentheses correspond to simulations conducted at a tenfold reduced strain rate.}
    \label{tab:fit_sigma}
    \renewcommand{\arraystretch}{1.2}
    \begin{tabular}{|c||c|c|c|c|c|c|}
        \hline
        & \multicolumn{6}{c|} {BFDGE-DETDA} \\
        \hline
        $\epsilon$ & 0.03 & 0.05 & 0.125 & 0.25 & 0.375 & 0.5 \\
        \hline
        $\sigma_{xx}(\epsilon_x)$  / MPa & \num{29} & \num{45} & \num{104} & \num{119} (\num{125}) & \num{114} & \num{142} (\num{149}) \\
        $\sigma_{yy}(\epsilon_y)$ / MPa & \num{20} & \num{75} & \num{106} & \num{120} & --- & \num{76} \\
        $\sigma_{zz}(\epsilon_z)$ / MPa & \num{33} & \num{49} & \num{28} & \num{47} & --- & \num{152} \\
        $\overline{\sigma}(\epsilon)$ / MPa & \textbf{\num{27}} $\pm$ \num{7} & \textbf{\num{56}} $\pm$ \num{16} & \textbf{\num{79}} $\pm$ \num{45} & \textbf{\num{95}} $\pm$ \num{42} & \textbf{\num{114}} $\pm$ \num{0} & \textbf{\num{123}} $\pm$ \num{41} \\
        \hline
        \hline
        & \multicolumn{6}{c|} {DGEBA-DETA} \\
        \hline
        $\epsilon$ & 0.03 & 0.05 & 0.125 & 0.25 & 0.375 & 0.5 \\
        \hline
        $\sigma_{xx}(\epsilon_x)$  / MPa & \num{31} & \num{89} & \num{140} & \num{51} (\num{140}) & \num{154} & \num{204} (\num{285}) \\
        $\sigma_{yy}(\epsilon_y)$ / MPa & \num{71} & \num{83} & \num{119} & \num{104} & --- & \num{323} \\
        $\sigma_{zz}(\epsilon_z)$ / MPa & \num{43} & \num{87} & \num{86} & \num{171} & --- & \num{227} \\
        $\overline{\sigma}(\epsilon)$ / MPa & \textbf{\num{48}} $\pm$ \num{21} & \textbf{\num{86}} $\pm$ \num{3} & \textbf{\num{115}} $\pm$ \num{27} & \textbf{\num{109}} $\pm$ \num{60} & \textbf{\num{154}} $\pm$ \num{0} & \textbf{\num{251}} $\pm$ \num{63} \\
        \hline
    \end{tabular}
\end{table}

A consistent correlation was observed between the elastic moduli of each network and their stress responses under load. Higher stress levels under tension were sustained by the DGEBA-DETA system, and slower relaxation was exhibited compared to BFDGE-DETDA, consistent with its greater stiffness and cross-link density.

Overall, the agreement between simulated and experimental properties, across equilibrium, dynamic, and viscoelastic regimes, confirms that the epoxy models used in this study provide a reliable basis for the vibrational analyses presented in the main text.

\section{SAGPR model}
\label{sec:sagpr_models}

\section{Experimental Vibrational Shifts}
\label{sec:experimental}

\begin{figure}[htb!]
\centering
\includegraphics[width=1.0\textwidth]{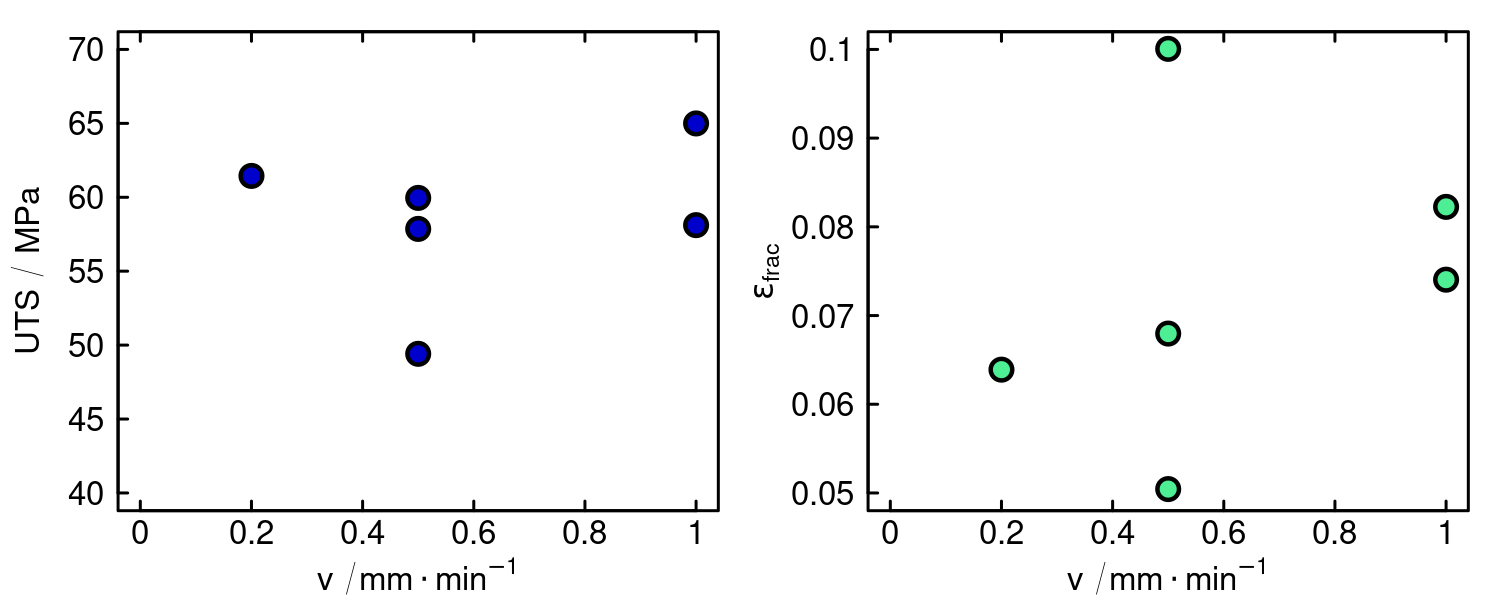}
\caption{Mechanical properties of epoxy films as a function of crosshead speed. Left: ultimate tensile strength (UTS). Right: strain at break ($\epsilon_\mathrm{frac}$). No systematic strain-rate dependence was observed.}
\label{fig:mech}
\end{figure}

To assess whether the deformation-induced vibrational response depends on the applied strain rate, we evaluated six independent tensile tests at crosshead speeds between \SI{0.2}{\milli\metre\per\minute} and \SI{1.0}{\milli\metre\per\minute}. The mechanical properties (Figure~\ref{fig:mech}) showed no systematic strain-rate dependence in terms of ultimate tensile strength or strain at break, with variability attributed to differences in film thickness and microscopic defects introduced during sample preparation.

\begin{figure}[htb!]
\centering
\includegraphics[width=1.0\textwidth]{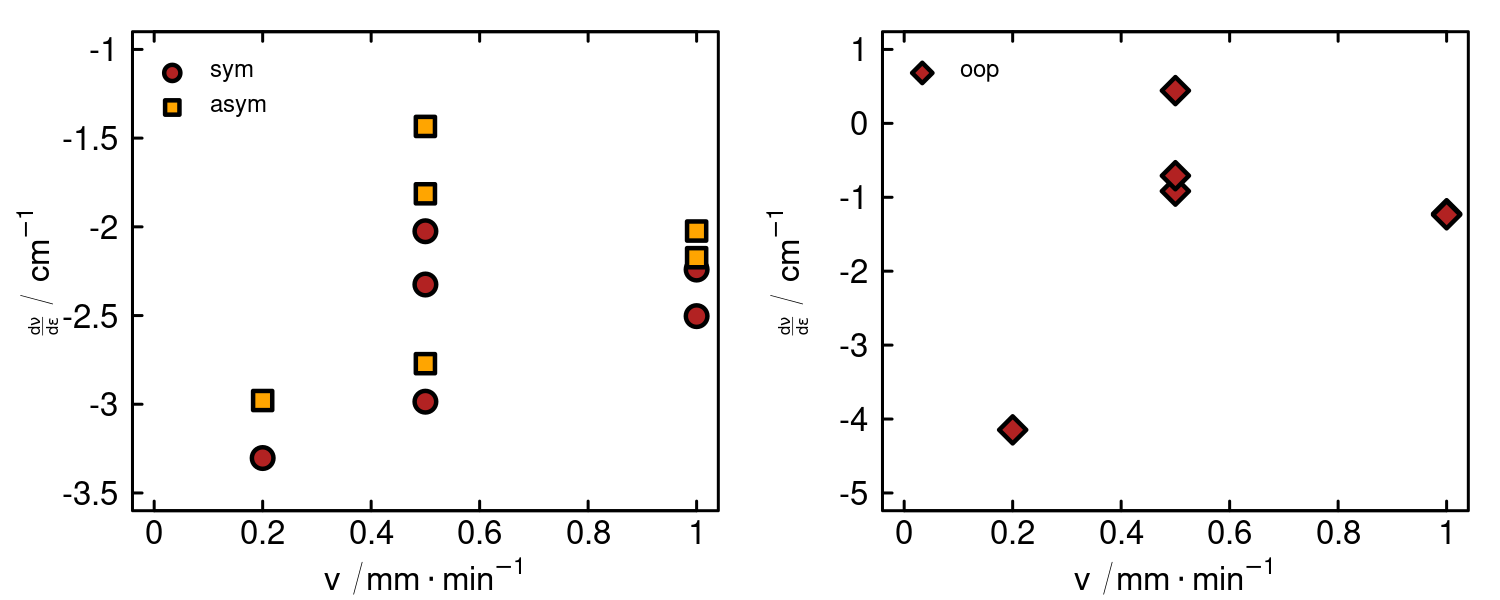}
\caption{Slopes of strain-dependent peak shifts ($\mathrm{d}\nu/\mathrm{d}\epsilon$) as a function of crosshead speed. Left: symmetric ($\nu_\mathrm{C_1-C_4}^{\mathrm{sym}}$) and asymmetric ($\nu_\mathrm{C_1-C_4}^{\mathrm{asym}}$) stretches. Right: out-of-plane bending vibration ($\nu_\mathrm{C-H}^{\mathrm{oop}}$). The stretches show a tendency toward larger redshifts at lower strain rates, while the oop mode remains scattered.}
\label{fig:peak_strainrate}
\end{figure}

The corresponding slopes of the strain-dependent vibrational shifts are shown in Figure~\ref{fig:peak_strainrate}. Both the $\nu_\mathrm{C_1-C_4}^{\mathrm{sym}}$ and $\nu_\mathrm{C_1-C_4}^{\mathrm{asym}}$ stretching modes display consistent linear redshifts, whereas the $\nu_\mathrm{C-H}^{\mathrm{oop}}$ bending vibration remains scattered without a reproducible trend. Although the dataset is not exhaustive, a tendency emerges: at slower strain rates, the stretching vibrations exhibit slightly larger redshift slopes. This behavior may reflect the longer time available for polymer backbones to align with the tensile axis, thereby amplifying the vibrational response to strain.

\begin{figure}[htb!]
\centering
\includegraphics[width=1.0\textwidth]{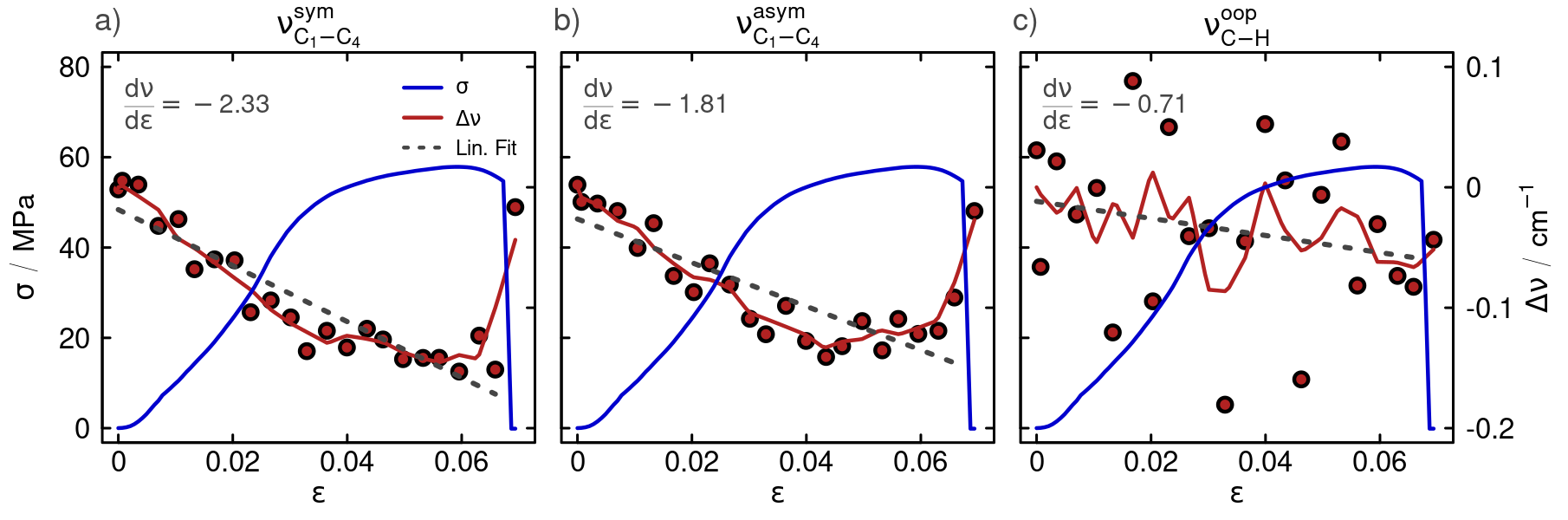}
\caption{Experimentally measured peak shifts during tensile deformation at a reduced strain rate of \SI{0.2}{\milli\metre\per\minute}. a) $\nu_\mathrm{C_1-C_4}^{\mathrm{sym}}$ stretch, b) $\nu_\mathrm{C_1-C_4}^{\mathrm{asym}}$ stretch, c) $\nu_\mathrm{C-H}^{\mathrm{oop}}$ bending vibration. Blue: stress–strain response ($\sigma$); red: peak shift ($\Delta\nu$); black circles: fitted peak positions; dashed gray lines: linear fits.}
\label{fig:peaks_exp_slow}
\end{figure}

To further evaluate the robustness of the strain–frequency correlation, we explicitly examined the case of a reduced crosshead speed of \SI{0.2}{\milli\metre\per\minute}. The strain-dependent frequency shifts are shown in Figure~\ref{fig:peaks_exp_slow}. Both $\nu_\mathrm{C_1-C_4}^{\mathrm{sym}}$ and $\nu_\mathrm{C_1-C_4}^{\mathrm{asym}}$ stretching modes again display clear linear redshifts that are consistent with the simulation results and the data presented in the main manuscript. In contrast, the $\nu_\mathrm{C-H}^{\mathrm{oop}}$ bending vibration shows no systematic trend and, in this particular sample, does not exhibit the blueshift that was observed in the representative dataset of the main text. This highlights that while the stretching modes provide robust markers of strain, the out-of-plane vibration is more sensitive to experimental conditions and less reliable as a quantitative probe.

\begin{figure}[htb!]
    \centering
    \includegraphics[width=0.5\textwidth]{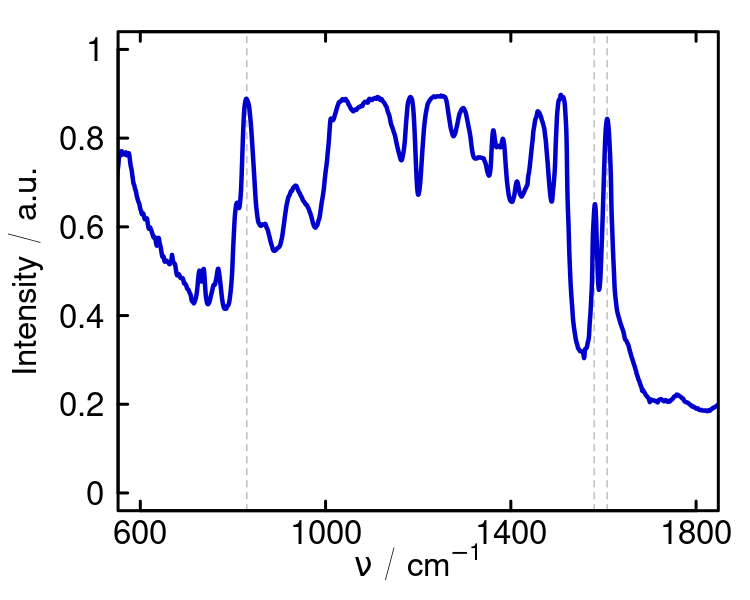}
    \caption{Reference transmission spectrum of the undeformed epoxy film in the mid-infrared range. Vertical dashed lines mark the regions associated with the $\nu_{\mathrm{C}-\mathrm{H}}$ bending and $\nu_{\mathrm{C}_1-\mathrm{C}_4}$ stretching vibrations.}
    \label{fig:spectrum}
    \end{figure}

For completeness, Figure~\ref{fig:spectrum} shows the reference infrared spectrum of the undeformed epoxy film in the range of 600–1800\,cm$^{-1}$. The data is presented as $1-T$, i.e. the inverse transmission, to enhance the visibility of vibrational absorption bands. The highlighted regions indicate the vibrational modes analyzed under mechanical load.

\bibliographystyle{ieeetr} 
\bibliography{ref.bib}